\documentclass[pra,twocolumn,amsfonts,superscriptaddress,a4paper,showpacs,showkeys]{revtex4}
\usepackage{graphicx}
\newcommand{\gb}{\beta}
\newcommand{\gd}{\delta}

\newcommand {\sg} {\sigma}

\begin{document}
\date{\today}
\title[Short Title]
{Critical dynamics and universality of the random-bond Potts ~\\
ferromagnet with tri-distributed quenched disorders
~  }
\author{H.P. Ying}
\affiliation{Zhejiang Institute of Modern Physics, 
Zhejiang University, Hangzhou 310027, P.R. China}
\affiliation{Fachbereich Physik, Universit\"at Siegen, 
D-57068 Siegen, Germany }
\author{B.J. Bian}
\affiliation{Department of Mathematics, 
Zhejiang University, Hangzhou 310027, P.R. China \\~\\}
\author{D.R. Ji$^1$}
\author{H.J. Luo$^2$}
\author{ L. Sch\"ulke$^2$}

\vspace{0.5in}
\begin{abstract} 
Critical behavior in short-time dynamics is investigated by a Monte Carlo study 
for the random-bond Potts ferromagnet with a trinary distribution 
of quenched disorders on two-dimensional triangular lattices. 
The universal dynamic scaling is verified and applied to estimate the
critical exponents $\theta$, $z$ and $\beta/\nu$ for several realizations 
of the quenched disorder distribution. 
Our critical scaling analysis strongly indicates that the bond randomness
influences the critical universality. 
\end{abstract} 


\vspace{0.2in}

\maketitle

%%%%%%%%%%%%%%%%%%%%%%%%%%%%%%%%%%%%%%%%%% 
\section{Introduction } 
%\paragraph{\bf Introduction} 
The study of the critical properties of physical systems with bond 
randomness on phase transitions is a quite active field of current 
interest in equilibrium statistical physics \cite{Cardy96,Cardy99}.
One of the central importance here is to answer the question whether
the critical exponents (of a homogeneous pure magnet) change on
the addition of quenched impurities and, if so, how do they change?
Replying to this question, it was first stated already two decades ago that 
if the specific heat critical exponent $\alpha$ of a pure system is 
positive, then a quenched disorder is a relevant perturbation at the 
second-order critical point and it causes changes in critical 
exponents. This statement is known as the Harris criterion \cite{Harris}.
Furthermore, following the earlier work of Imry and Wortis \cite{Imry79}, 
who argued that quenched disorder could smooth {\it first-order} 
phase transitions and thus produce the {\it second-order}
phase transitions, the introduction of randomness to pure systems 
undergoing a first-order transition has been comprehensively considered 
\cite{Hui89,Aizen89}.
The theory was initially numerical checked with the Monte Carlo (MC) 
method by Chen, Ferrenberg and Landau (CFL) \cite{Chen92}, who studied the 
8-state random-bond Potts model with a bimodal self-dual distribution.
On the other hand the experimental evidence has been found that, for the 
order-disorder phase transitions of absorbed atomic layers on 
two-dimensions, the critical exponents are changed from the original 
universality class of the 4-state Potts model on the addition of 
disorders \cite{Schwen94,Voges98}. 
However, no modification is
found when the pure system belongs to the Ising universality class
\cite{Mohan98}. Recently such disordered systems are extensively
studied by examining how a phase transition is modified
by quenched disorder coupling to the local energy density \cite{Folk00},
where use of intensive MC simulations is often helpful 
\cite{Wise95,Kardar95,Yasar98,Chate98a,Olson99}.  

The two-dimensional (2D) $q$-state random-bond Potts ferromagnet (RBPF) 
is an interesting framework in the MC research to study the influence 
of impurity on pure systems. 
For $q > 2$ such randomness acts as a relevant perturbation, and for 
$q > 4$ it even changes the nature of the transition from first to second 
order. In Table 1 we list the magnetic scaling index 
$\beta/\nu$ of the 2D 8-state RBPF obtained by different groups. 
Here a disorder amplitude $r$, defined by the ratio
of the {\it strong} to {\it weak} coupling (distributed according to the 
bimodal distribution) in the range $r=$2-20 appears to be adapted to 
a numerical analysis and gives a good estimate of the disordered fixed
point exponents \cite{Cardy99,Picco98}. A recent work of Olson and Young 
\cite{Olson99} used a specially {\it continuous} self-dual probability 
distribution of the disoreded bonds 
\begin{eqnarray}
P_X(x) =\frac{2\sqrt{q}}{\pi(1-x)^2+qx^2}~~~~~
\label{OY}
\end{eqnarray}
for the Boltzmann factor $x\equiv e^{-K}$, and performed a MC study of 
multiscaling properties of the correlation functions for several values 
of $q$. Their results are very interesting to examine the universality 
class of the RBPF. 
Cardy and Jacobsen \cite{Cardy97} studied the RBPF based on the connectivity 
transfer matrix (TM) formalism \cite{Jaco98}, and their estimates of the 
critical exponents lead to a continuous variation of $\gb/\nu$ with $q$, 
which is in sharp disagreement with the MC results by CFL(see Table 1).
%Tab1
%%%%%%%%%%%%%%%%%%%%%%%%%%%%%%%%%%%%%%%%%%%%%%%%%%%%%%%%%%%%%%%%%%%%%%%%%%
\begin{table} \begin{center}
\vskip 0.1cm
\begin{tabular}{c  c  c  c } \hline\hline
      Authors      & $r$   &  $ \gb/\nu$ & Technique \\ \hline
    CFL\cite{Chen92}             & ~2,10~ & 0.118(2)  & MC \\ \hline
Chatelain and Berche\cite{Chate98a}& ~10~ & 0.152(3) & MC  \\ \hline
Olson and Young\cite{Olson99}    &$P_X(x)$& 0.156(3) & MC  \\ \hline
Picco\cite{Picco98}              &  ~10~  & 0.153(1) & MC  \\ \hline
Cardy and Jacobsen\cite{Cardy97} &  ~2~   & 0.142(4)  & TM \\ \hline
Chatelain and Berche\cite{Chate99} & ~10~ & 0.1505(3) & TM  \\ \hline
Ying and Harada\cite{Ying00}     &  ~10~  & 0.151(3) & STD \\ \hline\hline
\end{tabular}
\caption { Magnetic exponent $\gb/\nu$ estimated by different
groups for the 2D 8-state RBPF. }
\end{center} \end{table}
%%%%%%%%%%%%%%%%%%%%%%%%%%%%%%%%%%%%%%%%%%%%%%%%%%%%%%%%%%%%%%%%%%%%%%%%%

In this paper, we present a MC study by short-time dynamics (STD) 
to verify the dynamic scaling 
features of the RBPF and estimate the critical exponents for a
{\it trinary} random-bond Potts model on a 2D triangular lattice. It is 
well known that this model on the 2D triangular lattice belongs to the 
same universality as the one on the 2D square lattice in equilibrium. 
However, a difference 
exists that the transition temperature is changed because the systems 
on triangular lattices lost the original (square-square) self-duality 
relation. Instead, a honeycomb-triangular duality (or star-triangular 
duality) is satisfied \cite{Baxter}. Therefore the multi-disorder amplitudes 
$r_i$ can be introduced to study the dynamic universality of the RBPF 
on such lattices by the STD approach \cite{Janss89,Zheng98}. 
%\cite{Janss89,Huse89,Zheng98}. 
In particular we will investigate the 
dynamic behavior of critical scaling affected by introducing quenched 
randomness, and study the dependence of critical exponents
on the randomness strength in order to clarify the crossover behavior 
(pure $\longleftrightarrow$ random fixed point 
$\longleftrightarrow$ percolation-like limit) \cite{Picco98}.

\section{The model}
%\paragraph{\bf The model} 
The Hamiltonian of the $q$-state Potts model on a 2D triangular lattice 
with quenched random interactions can be written as
\begin{widetext}
\begin{eqnarray}
-\gb H=\sum_{<i,j>} K_{ij} \gd_{\sg_i \sg_j}
=\sum_{\{L_0\}}K_0\gd_{\sg_i\sg_j}+\sum_{\{L_1\}}K_1\gd_{\sg_i\sg_j}
+\sum_{\{L_2\}}K_2\gd_{\sg_i\sg_j},~
\label{Ham1}
\end{eqnarray}
\end{widetext}
where the spin $\sg_i$, defined on lattice site $i$, takes the values 1,
$\cdots ~ q$, $\gb=1/k_B T$ is the inverse temperature, $\gd$ the Kronecker 
{\it delta} function and the sum is over all the nearest-neighbor (NN) 
interactions (bonds) $<i,j>$ on the lattice with a size $N=L^2$. 
There are $3N$ NN bonds (six for each and shared by two spins)
and they can be grouped into three classes $\{L_0,L_1,L_2\}$
corresponding to the interactions on them. The dimensionless
couplings $K_{ij}$ can be selected from three positive
(ferromagnetic) values $K_0$, $K_1=r_1 K_0$ and $K_2=r_2K_0$  
according to a probability distribution,
\begin{widetext}
\begin{equation}  \label{eq2}
P(K)=(1-p_1-p_2)\gd (K -K_0)+p_1\gd (K -K_1)+p_2\gd (K -K_2).~~~~~~~~~~~~
\end{equation}
\end{widetext}
When $p_1=p_2=1/3$ it is a trinary random-bond system. 
By denoting $F_0=(e^{K_{c}}-1)$, $F_1=(e^{K'_{c}}-1)$ and $F_3=(e^{K''_{c}}-1)$,
where $K_c$, $K'_c=r_1K_c$ and $K''_c=r_2K_c$ are the critical values of $K_0$, 
$K_1$ and $K_2$ respectively, 
the critical point can be determined by a self-dual relation
\cite{Baxter,Kim74},
\begin{eqnarray}
F_0 F_1 F_2+F_0 F_1+F_0 F_2+F_1 F_2=q~~~~
\label{eq3}
\end{eqnarray}
for the trinary system with the given state parameter $q$.

The strong to weak coupling ratio $r_i=K_i/K_0$ is called 
{\it disorder amplitude}. The values $r_1=0$ or $r_2=0$ correspond 
to the diluted case \cite{Mar99}, 
and $r_1= r_2=1$ to the pure case where the phase transitions are 
first-order for $q > 4$. With the presence of quenched random-bonds, 
however, the {\it second-order} phase transitions are 
induced for any $q$-state of the Potts model and the critical points
can be calculated according to Eq.(\ref{eq3}).
%for different values of disorder amplitude $r_i$.
In our simulations we chose $q=8$ which is known to have a strong 
first-order phase transition in the pure cases, in hope that we would find 
second-order phase transitions induced by the quenched random disorders 
to demonstrate the influence of quenched impurities on the 
{\it first-order} phase transitions. The strength of disorder is realized 
by the disorder amplitudes $\{r_1,r_2\}$ as chosen in Table 2.
There are three types of such disordered systems for the trinary distribution:
(I) one-third bonds chosen randomly are strongly coupled with \{$r_1$=1, 
$r_2>1$\}, (II) two-third bonds strongly coupled with different amplitudes 
$r_2>r_1>1$ and (III) two-third bonds with the same amplitude $r_1$=$r_2>1$.
Actually the systems III, 2/3 bonds being strongly coupled with
\{$r_1$=$r_2>1$\}, are related to those I,
1/3 bonds being strongly coupled, % with \{$r_1$=1, $r'_2>1$\},
because the former can be transferred to the latter by taking 1/3 bonds 
to be {\it weakly} coupled with \{$r_1$=1, $r'_2=1/r_2<1$\}.
%%%%%%%%%%%%%%%%%%
The study will be concentrated on the important question whether there 
exists an Ising-like universality class for systems with 
multi-quenched randomness \cite{Kardar95,Chate99}.  
%was done in \cite{Chen92,Picco96,Chate99}, to check the Ising-like
%universality class.
%Tab2
%%%%%%%%%%%%%%%%%%%%%%%%%%%%%%%%%%%%%%%%%%%%%%%%%%%%%%%%%%%%%%%%%%%%%%%%%%
\begin{table} \begin{center}
\vskip 0.1cm
\begin{tabular}{ l |l |l }  \hline \hline
~I:  &~II: &~III \\ \hline
\{$r_1,r_2$\}~ $T_c$~&\{$r_1,r_2$\}~ $T_c$~&\{$r_1,r_2$\}~ $T_c$~\\ \hline
\{1, 5\} ~.39903934...&\{2, 5\} ~.33624310...&\{~5, 5\} ~.19047260...\\
\{1, 8\} ~.29663628...&\{5, 8\} ~.19047260...&\{~8, 8\} ~.15574523...\\
\{1,10\}~.25548102...&\{8,10\}~.13987343...&\{10,10\}~.12638846...\\ 
\{1,12\}~.22533080...&\{2,12\}~.19784059...&\{12,12\}~.11306096...\\ \hline \hline
%%%%%%%%%%%%%%%
\end{tabular}
\caption { The critical temperature as a function of disorder amplitudes 
$\{r_1$, $r_2\}$ for the trinary RBMF.
(I) 1/3 bonds are strongly coupled with $r_2>1$, 
(II) 2/3 bonds are strongly coupled with different amplitudes 
$r_2>r_1>1$ and (III) with the same amplitude $r_1$=$r_2>1$.  }
\end{center} \end{table}
%%%%%%%%%%%%%%%%%%%%%%%%%%%%%%%%%%%%%%%%%%%%%%%%%%%%%%%%%%%%%%%%%%%%%%%%%%

\section{Short-time dynamics}
%\paragraph{\bf Short-time dynamics} 
For a long time, it was believed that universality and scaling relations
can be found only in equilibrium or in the {\it long-time} regime.
In Ref.\cite{Janss89},  however it was discovered that for an $O(N)$ 
vector magnetic system in the states with a very high temperature 
$T \gg T_c$, when it is suddenly quenched to the critical temperature $T_c$ 
and evolves according to a dynamics of model A, a universal dynamic scaling 
behavior emerges already within the short-time regime,
\begin{equation}\label{eq4}
M^{(k)}(t,\tau,L,m_0)=b^{-k\gb/\nu}M^{(k)}(b^{-z}t,b^{1/\nu}\tau,
b^{-1}L,b^{x_0}m_0),
\end{equation}
where $M^{(k)}$ is the $k$th moment of magnetization, $t, \tau=(T-T_c)/T_c,
L$ and $b$ are time, reduced temperature, linear size of the lattice and
scaling factor respectively. $\gb$ and $\nu$ are the well known static
critical exponents. The quantity $x_0$, a {\it new independent} exponent, 
is the scaling dimension of the initial magnetization $m_0$. This dynamic 
scaling form is generalized from finite size scaling in equilibrium.

Up to present, the short-time dynamic MC simulations have been successfully 
performed to estimate the critical temperatures $T_c$ and the critical 
exponents $\theta$, $\gb$, $\nu$ and the dynamic exponent $z$ for the 
various models with second order phase transitions \cite{Zheng98}.
This approach has also been extensively applied for systems with
disorders, as the FFXY and spin glass, to estimate both 
dynamic and static critical exponents \cite{Luo98,Luo99},
and for systems on the 2D triangular lattices
to study the dynamic scaling universality \cite{Ying01}.
The method is also efficient to study the deterministic dynamics
\cite{Zheng99}.

We carry out simulations in an analogous strategy described in 
Ref.\cite{Ying00} where the magnetic observables to be measured for the
RBPF have been well defined. In this work, however, an additional step 
is introduced to verify that the {\it second-order} phase transitions 
really are induced when quenched randomness is added. Then we will 
be particularly concerned to demonstrate the evidence for 
the dynamic universality of the trinary RBPF. 
Therefore our MC simulations are mainly performed for 
the time evolution of the magnetization $M(t)$ from the disordered 
initial states with small magnetizations 
%$m_0\sim0$, 
$m_0\gtrsim0$, 
and the completely 
ordered state with $m_0=1$. In the former, the magnetization  $M(t)$
undergoes an initial increase at the critical point $K_c$ 
for $t > t_{mic}$,
\begin{eqnarray}
M(t) \sim m_0 t^\theta,~~~~~~~\theta = (x_0 - \gb/\nu)/z.~~~~
\label{eq5}
\end{eqnarray}
For the latter, with $m_0=1$, $M(t)$ shows a power-law 
decay behavior 
\cite{Ito93}
\begin{eqnarray}
M(t) \sim t^{-c_1} , ~~~~~c_1 =\gb/\nu z, ~~~~~~~~~
\label{eq6}
\end{eqnarray}
at the critical point.  Furthermore the Binder cumulant
$U(t,L)=M^{(2)}(t,L)/M^2(t,L)$ shows, for $m_0=1$ at criticality, 
a similar power-law behavior on a large enough lattice,
\begin{eqnarray}
 U(t,L) \sim t^{c_u},~~~ c_u = d/z.~~~~~
\label{eq7}
\end{eqnarray}
Here the magnetization $M(t)$ is defined by
\begin{eqnarray}
M(t)&=&\frac{1}{N}\left [\left < (\frac{qM_o(t) -N}{q-1})\right >\right ],~~~~~
\label{eq8}
\end{eqnarray}
$M_o =\mbox{max}(M_1, M_2, \cdots, M_q)$ with $M_i$ being the
number of spins in the $i$th state. 
$N=L^2$ is the total number of spins on the lattice.
$<\cdots>$ denotes thermal averages 
over the initial states and/or random number sequences, and 
$[\cdots]$ the disorder averages over quenched randomness distributions. 
The time unit $t$ is defined as a MC sweep over all spins on the lattice.

The time-dependent susceptibility (the second moment of the magnetization) 
is defined as usual
\begin{eqnarray}
M^{(2)}(t)&=&\frac{1}{N}\left [\left (<M^2(t)>-<M(t)>^2\right )\right ].~~~
\label{eq9}
\end{eqnarray}
It has played an important role in equilibrium to determine 
the critical exponents $\gamma/\nu$ and $\beta/\nu$ \cite{Chen92}.

\section{Results}
%\paragraph{\bf Results} 
In our simulations, up to 60,000 MC processes are taken for MC averages 
(300 samples chosen as disorder averages in a given distribution and 
100-200 initial configurations as thermal averages for each disorder 
configuration realization). Statistical errors are simply 
estimated by performing three groups of averages with different random 
number sequences as well as independent initial configurations.
To minimize the number of bond configurations needed for the
disorder averaging, we confine our study to the bond distributions in
which there are the same number of disordered (stronger) NN bonds for
each of the three groups $\{L_0,L_1,L_2\}$.
This procedure is reliable and should reduce the variation
between different disorder configurations with no loss of generality.
Periodic boundary conditions are applied to the $N=L^2$ lattice.
We use $L$ up to 128 and adopt the heat-bath algorithm.
%Fig1.2
%%%%%%%%%%%%%%%%%%%%%%%%%%%%%%%%%%%%%%%%%%%%%%%%%%%%%%%%%%%%%%%%%%%%%%%%%%%%
\begin{figure}[htbp!]\centering
\includegraphics*[width=8.40cm]{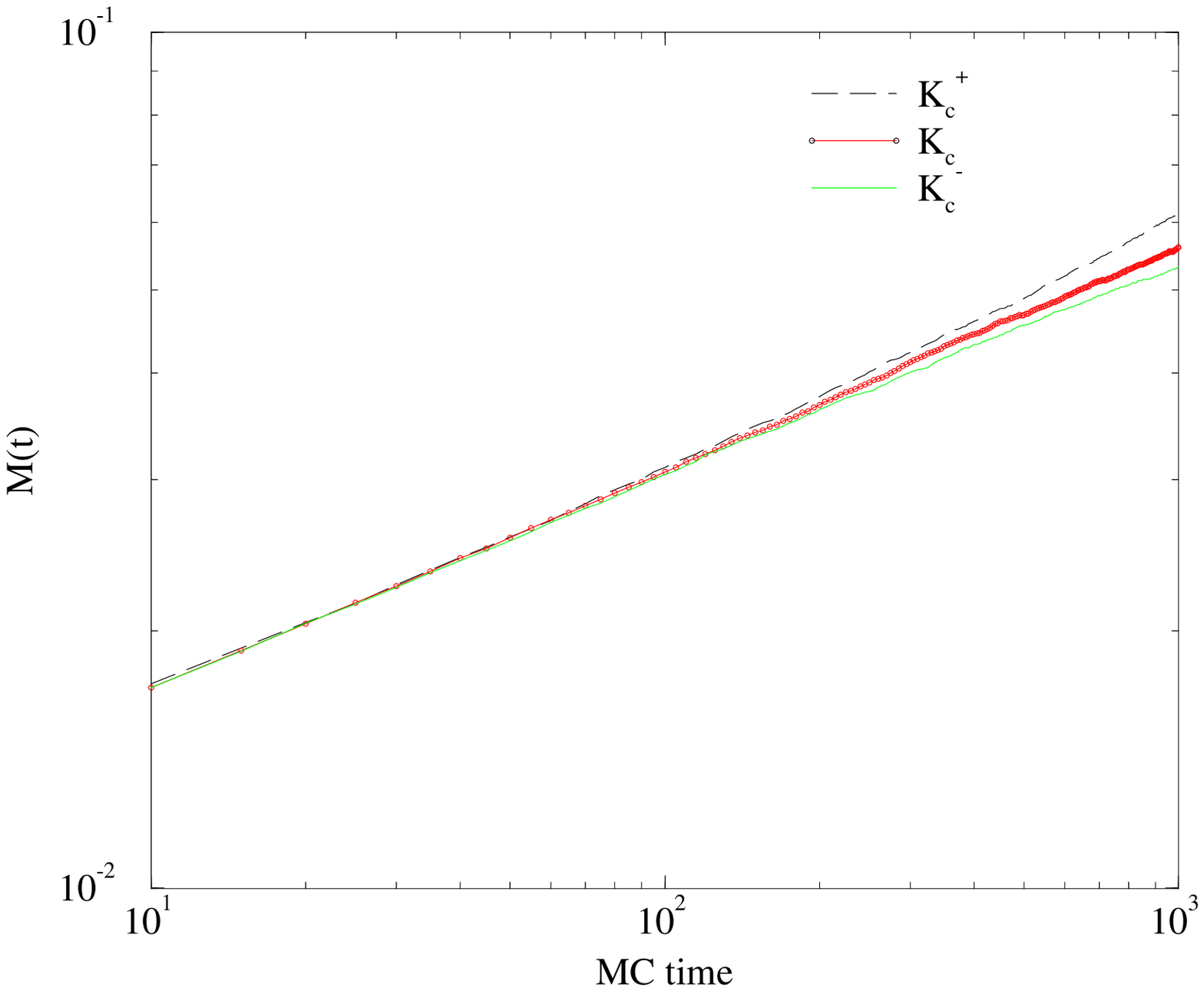}
\caption{$M(t)$ versus MC time plotted on a log-log scale for
\{$r_1$=1,$r_2$=8\} from the initial states $m_0\sim 0.01$, performed
at and near the critical point $K_c$ and $K_c^{\pm}=(1 \pm 0.012)K_c$. }
\label{Fig1}
%\end{figure}
%
%the curves should be in the form as shown in Figure~\ref{fig4}.
%
\vspace{0.20cm}
%\begin{figure}[htbp!]\centering
\includegraphics*[width=8.40cm]{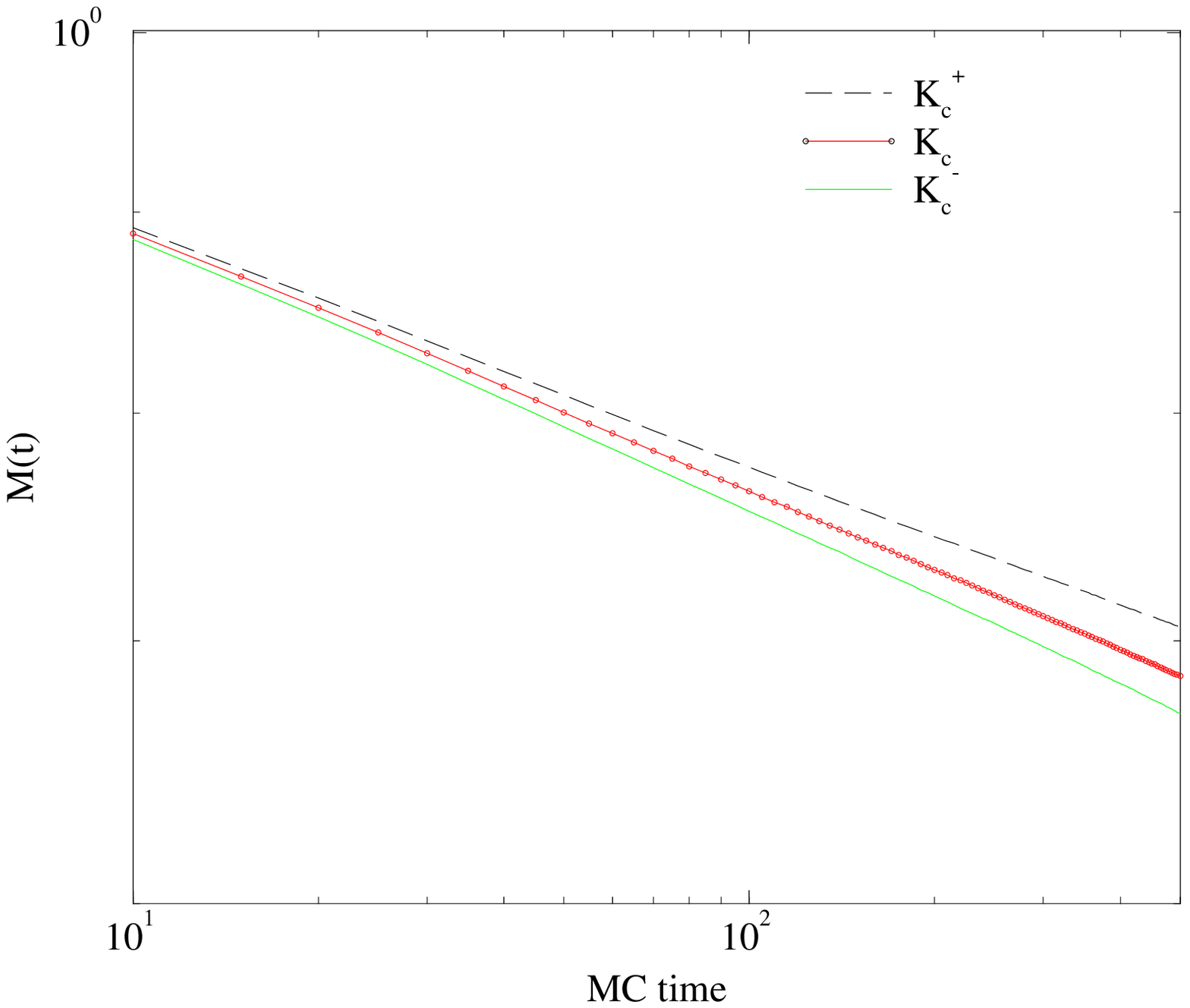}
\caption{The same as Fig.1 but from the ordered states $m_0=1$. ~~~~~~~}
\label{fig2}
\end{figure}
%%%%%%%%%%%%%%%%%%%%%%%%%%%%%%%%%%%%%%%%%%%%%%%%%%%%%%%%%%%%%%%%%%%%%%%%%%%%

The data for the evolution of $M(t)$ from 
$m_0\gtrsim 0$ 
and $m_0=1$ exhibit that the power-law behavior is satisfied at 
the {\it same} point $K_c$ for both 
the disordered and ordered initial states. This fact gives an evidence 
that a second-order phase transition is induced at $K_c$ instead of a 
first-order one: In Ref.\cite{Schu00} it was argued that there are two 
quasi-critical points if a first-order transition happens, $K^{random}_c$ 
for evolutions from random initial states and $K^{order}_c$ from ordered 
initial states, and $K^{random}_c < K_c < K^{order}_c$ is satisfied. 
We plot in Fig.1 and Fig.2 the curves $M(t)$ for 
disorder amplitudes \{$r_1$=1, $r_2=8$\} on a $64^2$ lattice. 
They exhibit obviously a power-law relaxation at 
the same critical point for both the 
$m_0\gtrsim0$ and $m_o=1$ states.
%$m_0\sim 0$ and $m_o=1$ states.
On the other hand the curves deviate from the power-law behavior for a small 
but nonzero $\tau$ described by $M(t,\tau)=m_0 t^\theta M(t^{1/\nu z}\tau)$ or
$M(t,\tau)=t^{-c_1} F(t^{1/\nu z} \tau)$ 
with respect to the disordered or ordered initial states respectively.
This modification has been initially used to determine the critical
temperatures \cite{Schu95,Jas99}.
The same features, as plotted in Figs. 1 and 2, exist for all 
other realizations of the randomness listed in Table 2. 
%Fig3.4
%%%%%%%%%%%%%%%%%%%%%%%%%%%%%%%%%%%
\begin{figure}[htbp!]\centering
\includegraphics*[width=8.40cm]{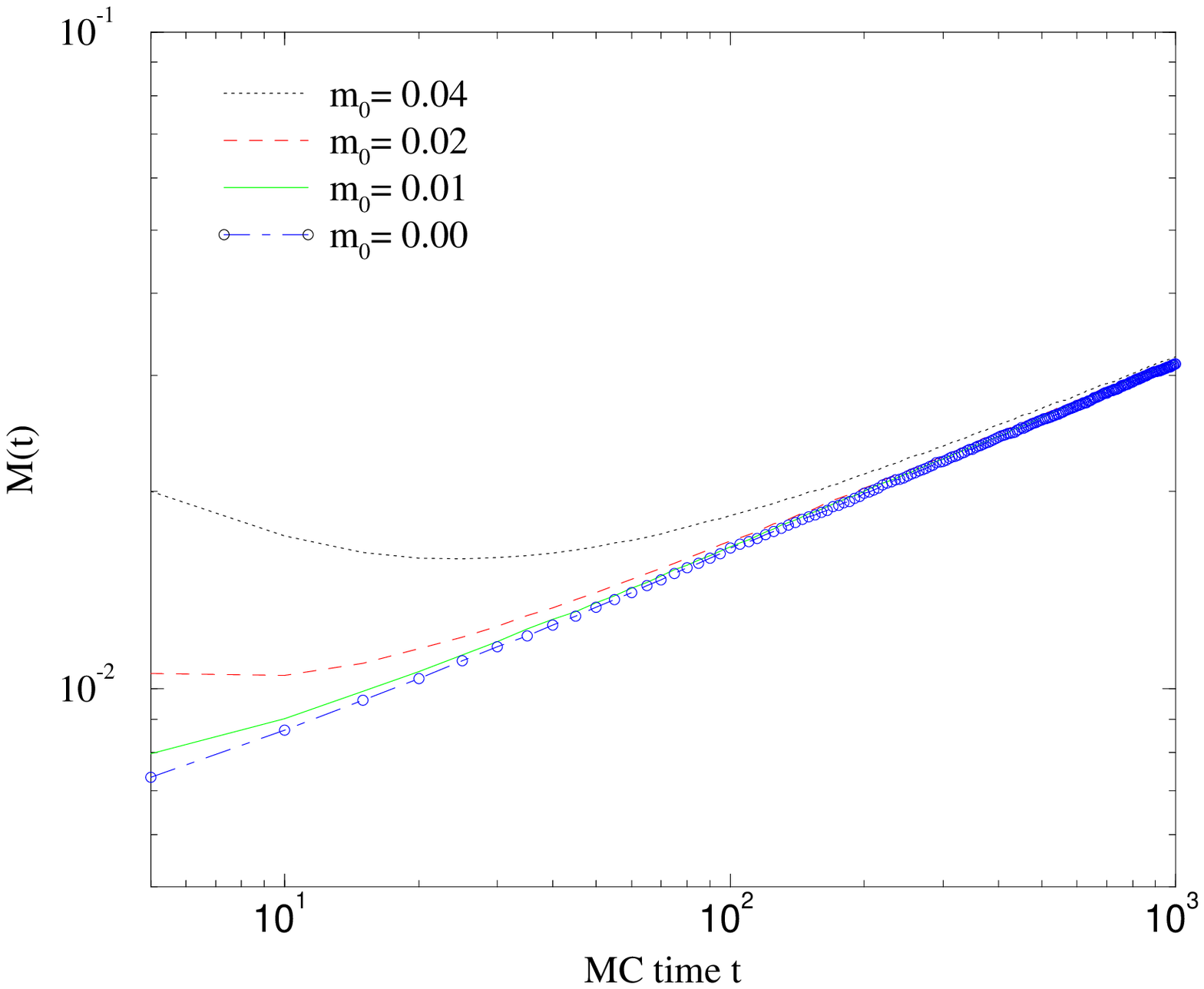}
\caption{$M(t)$ versus MC time plotted on a log-log scale for
\{$r_1$=5,$r_2$=8\} from random initial states with magnetizations
$m_0$=0.04,0.02,0.01 and with $m_0\sim 0$ on a $128^2$ lattice.  }
%\epsfbox{128r25m01-2-4-Mt.eps}
\label{fig3}
%\end{figure}
%
%the curves should be in the form as shown in Figure~\ref{fig4}.
%
\vspace{0.20cm}
%\begin{figure}[htbp!]\centering
\includegraphics*[width=8.40cm]{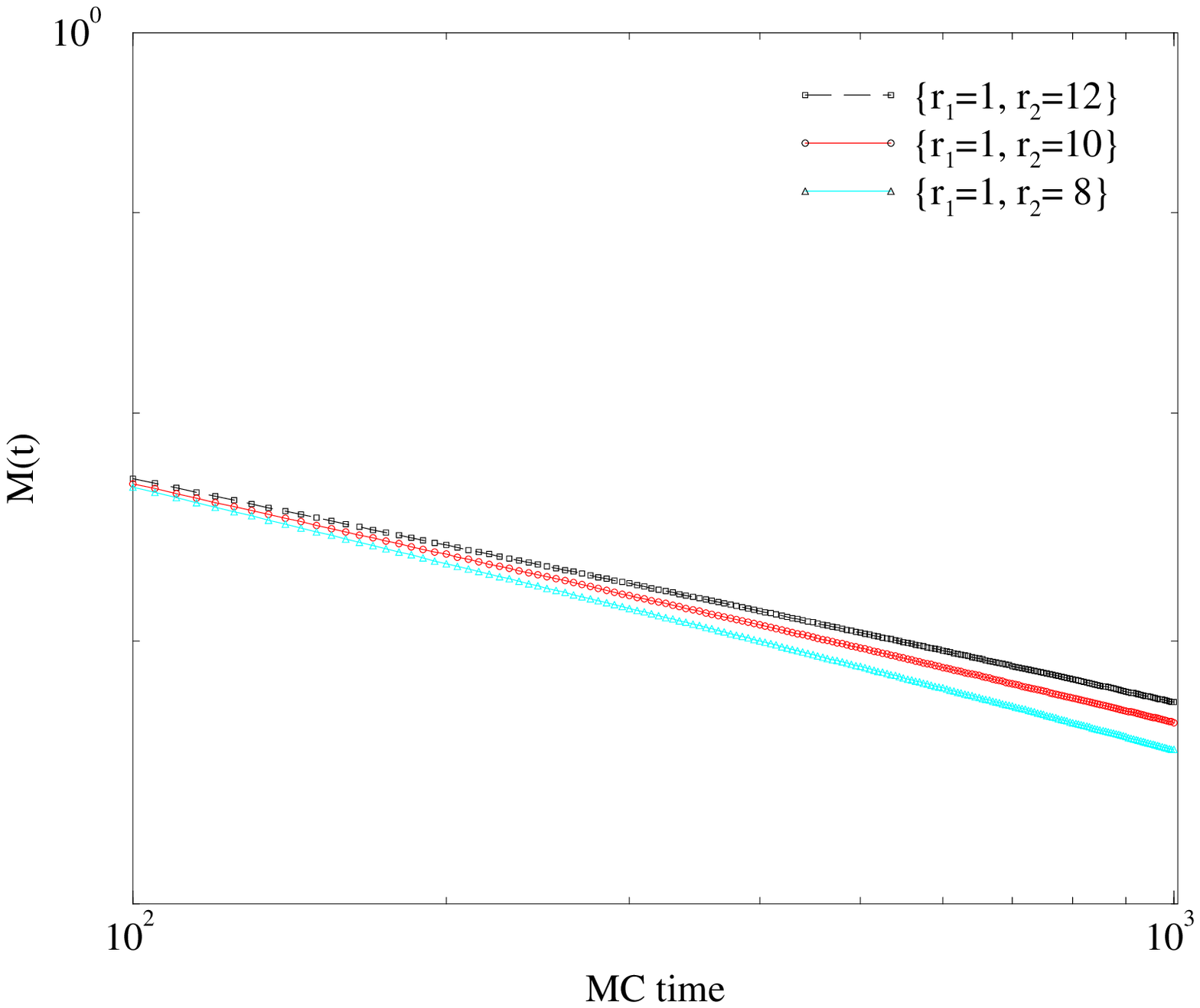}
\caption{$M(t)$ versus MC time plotted on a log-log scale for
\{$r_1$=1,$r_2$=8\}, \{$r_1$=1,$r_2$=10\} and \{$r_1$=1,$r_2$=12\}
from the initial states $m_0=1$ on a $128^2$ lattice.  }
\label{fig4}
\end{figure}
%%%%%%%%%%%%%%%%%%%%%%%%%%%%%%%%%%%

Then we perform simulations systematically at the exact critical points
$K_c(r_i)$ as a function of \{$r_1,r_2$\} for magnetization evolutions
starting with 
$m_0 \gtrsim 0$ and $m_0 =1$.
%$m_0 \sim 0$ and $m_0 =1$. 
The disordered initial states are prepared from random states with 
small magnetizations $m_0$ = 0.04 -- 0.01.
The curves $M(t)$ for the disordered configurations with $r_1=5$, $r_2=8$ on 
a $128^2$ lattice are displayed in Fig.3 with a double-logarithmic scale.
They exhibit the power-law behavior of Eq.(\ref{eq5}) for $t>t_{mic}$,
which depends on $m_0$. Thus the exponent
$\theta$ can be estimated with the standard least square fitting algorithm,
from the slopes of the curves for $t>t_{mic}$, and by
an extrapolation to the $m_0 \rightarrow 0$ limit.
The values are included in Table 3. 
Next we start from $m_0 =1$ to observe the power-law decay of
Eq.(\ref{eq6}) at the critical points $K_c(r_i)$. Some curves $M(t)$ are
plotted in Fig.4, which appear to have also a nice power-law behavior,
and the slopes are then used to estimate the index $c_1=\gb/\nu z$
(for results see Table 3).

Now we turn to the dynamic scaling behavior of the time-dependent Binder 
cumulant from $m_0 =1$ to estimate the scaling index $c_u=d/z$. 
Unlike the relaxation from the disordered state, 
the fluctuations involved in the ordered initial states
(e.g., $\sg_i$ = 1, $i=1,N$) are much smaller. 
Therefore, the measurements of the critical exponents based on 
Eq.(\ref{eq6}) and Eq.(\ref{eq7}) are better in quality than those 
from disordered states on Eq.(\ref{eq5}). So we take about 100 MC sweeps 
for random number sequences to be the thermal averages for the completely 
ordered initial state. 
In Figs.5 the curves $U(t)$ on a $128^2$ lattice are presented
and they exhibit a power-law behavior at the critical points as a function
of disorder amplitudes \{$r_1$, $r_2$\}.
Finally, based on the results of $c_u=d/z$ and $c_1=\gb/\nu z$, 
the critical exponents $z$ and $\gb/\nu$ can be calculated
consistently and their results are summarized in Table 3. 
%Tab3
%%%%%%%%%%%%%%%%%%%%%%%%%%%%%%%%%%%%%%%%%%%%%%%%%%%%%%%%%%%%%%%%%%%%%%%%%%
\begin{table} \begin{center}
\vskip 0.5cm
\begin{tabular}{lc|c| cc| cc} \hline\hline
~ & \{$r_1,r_2$\} & $\theta$&  $c_1$   &  $c_u$   &  $z$    & $\gb/\nu$\\ \hline

~~I:& \{~1,~5\}  & .238(3) & .0780(8) & .789(6)  & 2.53(4) & .202(5)  \\
  ~ & \{~1,~8\}  & .206(2) & .0654(7) & .611(6)  & 3.28(5) & .212(5)  \\
  ~ & \{~1,10\}  & .187(2) & .0597(7) & .599(6)  & 3.34(5) & .200(5)  \\
  ~ & \{~1,12\}  & .180(2) & .0559(7) & .554(6)  & 3.61(5) & .202(5)  \\
\hline
~II:& \{~2,~5\}  & .268(3) & .0864(9) & .889(6)  & 2.25(4) & .194(5)  \\
% ~ & \{~2,12\}  & .279(3) & .0852(9) & .892(6)  & 2.24(4) & .191(5)  \\
  ~ & \{~5,~8\}  & .283(3) & .0849(9) & .893(7)  & 2.24(4) & .190(4)  \\
  ~ & \{~8,10\}  & .290(3) & .0830(9) & .897(7)  & 2.23(4) & .185(4)  \\
\hline
III:& \{~5,~5\}  & .305(4) & .0784(8) & .959(7)  & 2.09(3) & .164(4)  \\
  ~ & \{~8,~8\}  & .296(3) & .0811(9) & .907(8)  & 2.21(4) & .179(4)  \\
  ~ & \{10,10\}  & .293(3) & .0815(9) & .898(8)  & 2.23(4) & .182(4)  \\
\hline\hline
\end{tabular}
\caption {
The values of indices ($\theta$,$c_1$,$c_u$), and results of
exponents ($z$, $\gb/\nu$) with respect to the disorder amplitudes
$\{r_1,r_2\}$.
 }
\end{center} \end{table}
%%%%%%%%%%%%%%%%%%%%%%%%%%%%%%%%%%%%%%%%%%%%%%%%%%%%%%%%%%%%%%%%%%%%%%%%%

\section{Summary and Conclusion}
%\paragraph{\bf Summary and Conclusion} 
We have, for the first time, introduced the trinary distribution for
the 2D RBPF with multi-disorder amplitudes to study its critical 
behavior by the STD method. 
In the MC simulation the dynamic scaling behavior is verified and the 
power-law behavior is used to estimate both the dynamic and static 
exponents as a function of the disorder amplitudes $\{r_1,r_2\}$. 
It is found that the values of exponents $\theta$ and $\gb/\nu$ 
vary continuously with the disorder amplitudes, and  
they violate the Ising-like universality. 
Furthermore the values of
dynamic exponent $z$ are larger than those for systems without disorders
and they increase with the strength of disorder amplitudes \{$r_1,r_2$\}
for the systems I: $r_1=1$, $r_2>1$ and III: $r_1$=$r_2>1$ in Table 3.

%Fig5,6
%%%%%%%%%%%%%%%%%%%%%%%%%%%%%%%%%%%%%%%%%%%%%%%%%%%%%%%%%%%%%%%%%%%%%%%%%%%%
\begin{figure}[htbp!]\centering
\includegraphics*[width=8.40cm]{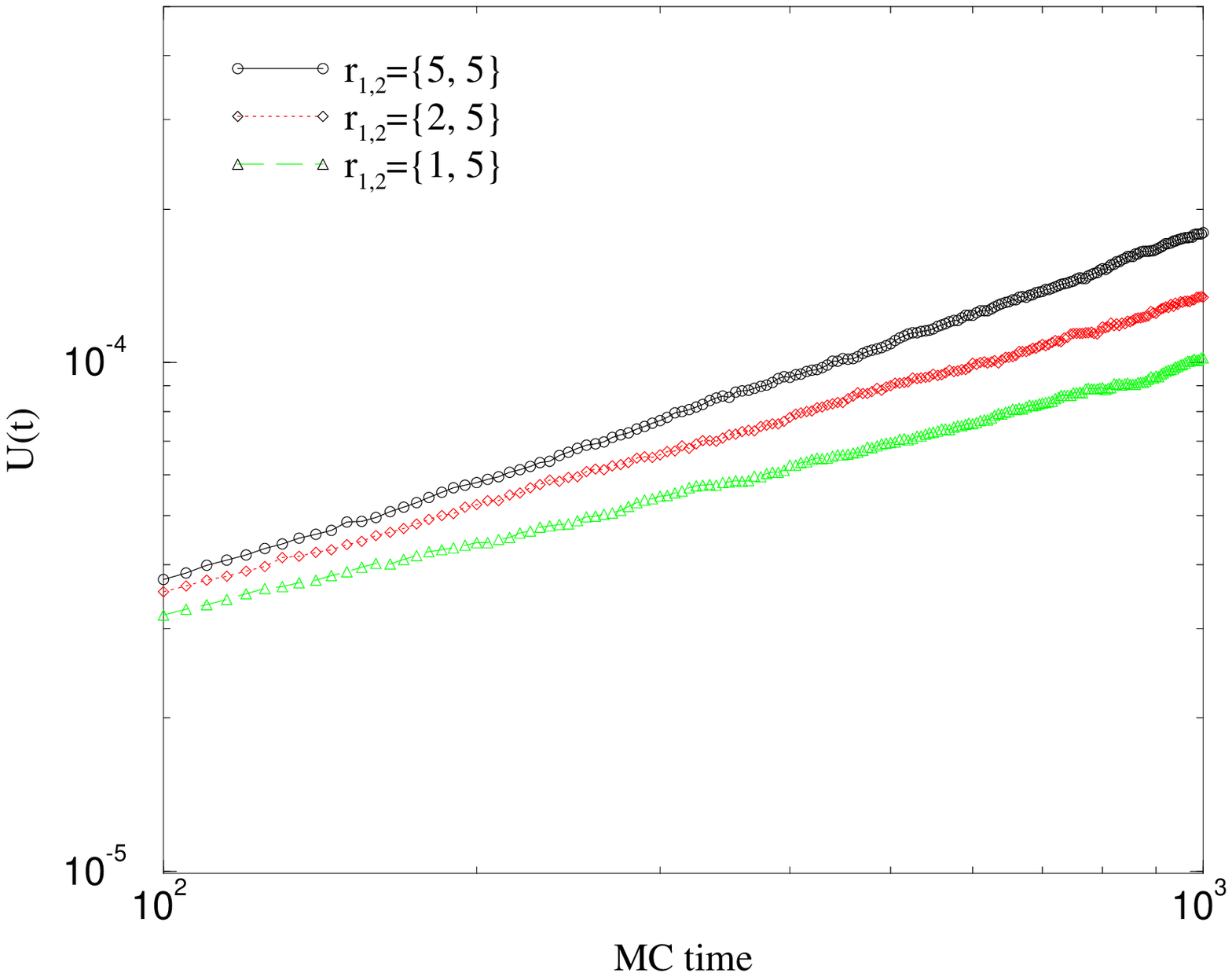}
\caption{$U(t)$ versus MC time on a log-log scale as a function
of the disorder amplitudes \{$r_1$, $r_2$\} from the initial state $m_0$=1,
used to estimate the index $c_u$ .}
%for \{$r_1$=1, $r_2$=5\}, \{$r_1$=2, $r_2$=5\} and \{$r_1$=5, $r_2$=5\}
\label{fig5}
%\end{figure}
%
%the curves should be in the form as shown in Figure~\ref{fig4}.
%
\vspace{0.20cm}
%\begin{figure}[htbp!]\centering
\includegraphics*[width=8.40cm]{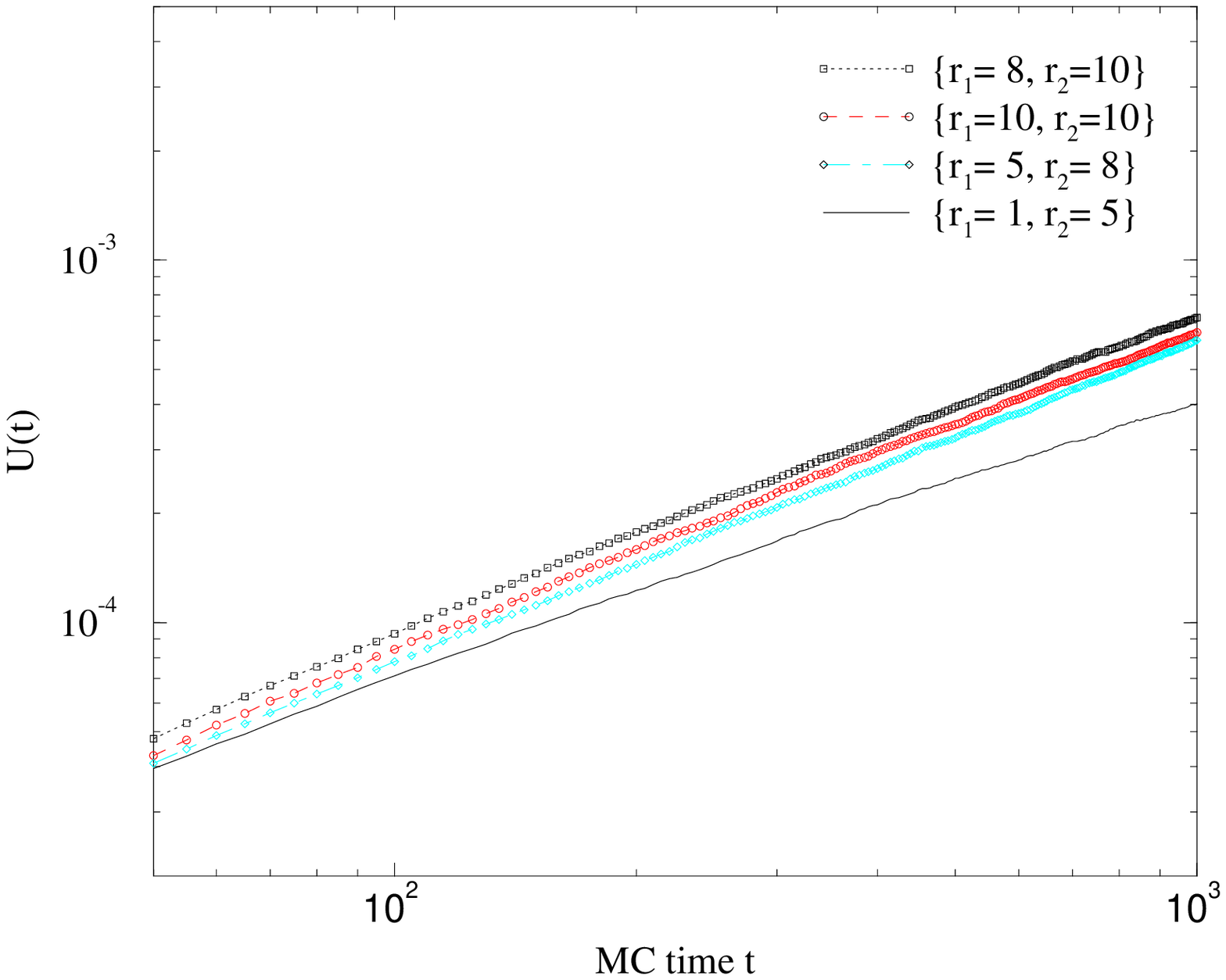}
\caption{The $U(t)$ curves for the disorder amplitudes \{$r_1$=5, $r_2$=8\},
\{$r_1$=8, $r_2$=10\} and \{$r_1$=10, $r_2$=10\} are almost parallel each other,
but it dose not for the \{$r_1$=1, $r_2$=5\} curve.
  }
\label{fig6}
\end{figure}
%%%%%%%%%%%%%%%%%%%%%%%%%%%%%%%%%%%%%%%%%%%%%%%%%%%%%%%%%%%%%%%%%%%%%%%%%%%%
On the other hand, for the systems II where 2/3 bonds are
strongly coupled with different amplitudes $1<r_1<r_2$,
they seem to have been located at a ``random'' regime where their
values of the critical exponents $\gb/\nu$ and $z$ are nearly independent
of disorder amplitudes within the error bars,
$z\sim 2.23$ and $\gb/\nu\sim 1.85$.
By comparing these results to $z$=2.23(4) and $\gb/\nu$=0.182(5)
for the $r_1$=$r_2$=10 in III,
we could argue that, after $r_1=r_2\geq 10$ (equivalent to
$r_1$=1, $r_2\leq 0.1$ in I) it will pass the ``crossover'' region
to the random regime from other type, I or III, of the disordered systems,
so that $z$ and $\gb/\nu$ will be nearly constants.
In Fig.6 the $U(t)$ curves for the disorder amplitudes 
\{$r_1=5,r_2=8$\}, \{$r_1=8,r_2=10$\} and \{$r_1=10,r_2=10$\} are almost 
parallel each other, which shows that the values of exponent $\gb/\nu$ are 
nearly same for these disordered systems.
%%%%%%%

In conclusion, our simulation verifies that second 
order phase transitions are induced. Then the results show 
evidence that the dynamic universality class of the trinary RBPF 
should not belong to that 
of the Ising model, as inferred by Jacobsen and Cardy \cite{Jaco98}.
From the work we find it rather encouraging to apply 
the short-time dynamic MC to simulate the scaling and critical 
dynamics of disordered spin systems. We will pursue this field 
to explore the logarithmical slow dynamics \cite{Cardy99,Ying01a}.

%\section{Acknowledgement} 
\vspace{0.1in}
{\it Acknowledgement}: 
%\paragraph{\bf Acknowledgement:} 
We are grateful to the helpful discussions with Q. Wang.
H.P.Y would like to thank the Heinrich-Hertz-Stiftung for fellowship 
and acknowledge the hospitality of Universit\"at Siegen.
%where the MC simulations were performed.
The work supported in part by the Deutsche Forschungsgemeinschaft,
DFG Schu 95/9-3 and by the NNSF of China, 19975041 and 10074055.

%%%%%%%%%%%%%%%%%%%%%%%%%%%%%%%%%%%%%%%%%%%%%%%%%%%%%%%%%%%%%%%%%%%%%%%%%%%%%
%\baselineskip=16.0pt

\end{document}